\documentclass[11pt]{article}
\usepackage{epsfig}
\usepackage{graphicx}
\usepackage{amssymb}      

\evensidemargin=\oddsidemargin  

\newcommand{\beq}{\begin{equation}}
\newcommand{\eeq}{\end{equation}}
\newcommand{\bq}{\begin{equation}}
\newcommand{\eq}{\end{equation}}
\newcommand{\ba}{\begin{array}}
\newcommand{\ea}{\end{array}}
\newcommand{\beqa}{\begin{eqnarray}}
\newcommand{\eeqa}{\end{eqnarray}}

\def\noi{\noindent}
\def\End{\end{document}}

\def\[{\left[}
\def\]{\right]}
\def\({\left(}
\def\){\right)}

\def\ra{\rightarrow}

\def\U1EM{U(1)_{\rm em}}

\def\[{\left[}
\def\]{\right]}

\begin{document}
\setcounter{footnote}{1}   
\input epsf

\def\thepage {}        

\begin{center}
 {\bf $\tau\ra\mu\ \eta$ in Supersymmetric Models }
\end{center}
\vskip 1.0cm
\begin{center}
    { {\sc Marc Sher}\\ Nuclear and Particle Theory Group,\\ College of
William and Mary, Williamsburg, Virginia 23187\\ \today}

\end{center}
\date{(~\today  \,~and~ hep-ph/0207136~)}  

\vspace{24pt}

\begin{abstract}
\noindent  The existence of large $\nu_\mu-\nu_\tau$ mixing suggests
the  likelihood of large smuon-stau mixing in supersymmetric models, 
leading to  $\mu$ and $\tau$ number violation.  In addition to 
interesting signatures in slepton and  neutralino production and decay,
this will lead to rare $\tau$  decays, such as $\tau\ra\mu\gamma$.  
Recently, it has been pointed  out that the $\tau\ra 3\mu$ branching
ratio could be substantial in  the large $\tan\beta$ region of
parameter space, due to an induced 
$\mu-\tau-$Higgs vertex.
  In this paper, another signature, $\tau\ra\mu\eta$ is considered.  In
the large 
$\tan\beta$ region, it is shown that the branching ratio of
$\tau\ra\mu\eta$ is 8.4  times the branching ratio of $\tau\ra 3\mu$,
independent of any  unknown parameters, and it will thus give the most
stringent  bound on Higgs-mediated lepton flavor violation, and may
provide its first signature.  In the other  regions of parameter space,
where $\tau\ra\mu\gamma$ is the most  prominent decay, the branching
ratio for $\tau\ra\mu\eta$ is  always substantially lower.

\end{abstract}

\noi The flavor physics of quarks and leptons is one of the most
prominent mysteries in  particle physics.  The most  surprising
development in flavor physics in the past decade has been  the
observation\cite{atmospheric} of very large mixing between muon  and
tau neutrinos.   The mixing will, at some level, lead to mixing  in the
charged lepton sector, giving violations of muon and tau  lepton number
conservation.

In the Standard Model supplemented with right-handed neutrinos,  such
violation is generally small\cite{pil}.  However, a  much
bigger effect will occur in supersymmetric models, by inducing  mixing
between the scalar muon and the scalar tau.   In the most  general
supersymmetric standard model, even  if neutrino mixing were not
present, one would have large mixing  between all scalar leptons and
scalar quarks.  This would lead to very  large flavor (quark and
lepton) changing neutral currents, which are  not observed.  It is thus
generally assumed that the squark and  slepton masses are equal at the
unification scale; an assumption that  occurs naturally in many models,
such as supergravity or  gauge-mediated supersymmetry breaking.

Even if the masses of the smuon and stau are equal at the unification 
scale, without mixing, the presence of non-diagonal neutrino mass 
terms (either from different Dirac mass terms or non-diagonal 
right-handed neutrino mass terms) will affect the masses through 
renormalization group running, and will generate mixing
terms\cite{bm}.  In  this paper, we will consider only mixing between
left-handed smuons  and staus; in most models, such mixing is the
largest.    It should be kept in mind that solar neutrino
oscillation experiments\cite{sno} also indicate large mixing between
muon and electron neutrinos, and in models with inverted hierarchies,
there could be substantial mixing between left-handed smuons and
selectrons, although the very strong bounds on $\mu\ra e\gamma$ will
constrain these effects.

The mixing is characterized by the parameter $\delta_{23}\equiv 
M_{23}^{2}/\tilde{m}^{2}$, where $M_{23}$ is the off-diagonal term in 
the slepton mass matrix and $\tilde{m}$ is the slepton mass scale.  
The value of $\delta_{23}$ is extremely model-dependent, of course, 
but in models in which the mixing arises entirely through 
renormalization group running\cite{babu,ellis,hinchliffe}, its value 
is typically between $0.1$ and $1$.

The most studied tau-number and muon-number violating process is 
$\tau\ra\mu\gamma$\cite{lots}.  Many of these works consider various 
models for $\delta_{23}$.  Normalizing the rate to the current 
bound\cite{nir,older}
\begin{equation}
    BR(\tau\ra\mu\gamma)= 1.1\times 10^{-6}\left({\delta_{23}\over 
    1.4}\right)^{2}\left({100\ {\rm GeV}\over \tilde{m}}\right)^{2}
    \end{equation} In the constrained MSSM, where the parameter space
is restricted to a  manageable dimensionality, this process dominates
in the low $m_{1/2}$  region\cite{ellis}.

In addition, one can consider tau-number and muon-number violation in 
production and decay of sleptons, neutralinos and 
charginos\cite{ellis,kalinowski,hisano}.  As an example, the process 
$\chi_{2}\ra\chi_{1}\mu\tau$ where $\chi_{1,2}$ are neutralinos, can be 
searched for at the LHC.  In the constrained MSSM, this process 
dominates\cite{ellis} in the $m_{1/2}>m_{0}$ region of 
parameter-space, and is thus complementary to $\tau\ra\mu\gamma$.

Recently, Babu and Kolda\cite{babu} pointed out that $\tau\ra 3\mu$ 
was a promising signature in models with a large value of 
$\tan\beta$.  Earlier\cite{babufirst}, they had noticed that squark
mixing would  induce a flavor non-diagonal quark-quark-Higgs Yukawa
coupling, and  had examined the consequences for rare B-decays.  The
same mechanism,  however, will also induce a $\mu-\tau-{\rm Higgs}$
vertex, and thus directly  to $\tau\ra 3\mu$, through tree-level Higgs
exchange (either the $h$, 
$H$, or $A$).  The branching ratio, for a reasonable choice of  mass
parameters, is 
$${\rm BR}(\tau\ra 3\mu)= (1 \times 10^{-7})\times\left({\tan\beta\over 
60}\right)^{6}\times \left({100\ {\rm GeV}\over m_{A}}\right)^{4}$$
where $m_{A}$ is the pseudoscalar mass.  This result is very 
insensitive to the SUSY spectrum, with the exception that it can 
increase by up to a factor of four for large $\mu$.  This branching 
ratio will be accessible at B-factories in the near future.

In this paper, another signature of lepton-number violation is 
discussed:  $\tau\ra\mu\eta$.    It will be shown that, in the large 
$\tan\beta$ region discussed in the previous paragraph, the branching 
ratio is much higher than $\tau\ra 3\mu$, and may be a much more 
sensitive test of Higgs-induced lepton flavor violation.   

In the Babu-Kolda model\cite{babu}, the lepton flavor violating 
interaction is given by
\begin{equation}
    {\cal L}_{LFV}=(2G^{2}_{F})^{1/4}{m_{\tau}\kappa_{32}\over 
    \cos^2\beta} 
    (\overline{\tau}_{R}\mu_{L})[H^{0}+iA^{0}]+{\rm h.c.}
    \end{equation} where $H^{0}$ and $A^{0}$ are the heavier scalar and
the pseudoscalar,  respectively, and we have chosen the generally
preferred region of  parameter-space in which $\sin(\alpha-\beta)\sim
1$.   Here, 
$\kappa_{32}$ is given in Ref. \cite{babu} and depends on loop integrals
(but is relatively  insensitive to SUSY parameters).   For SUSY
parameters $\mu=M_1=m_2=m_{\tilde l}=m_{\tilde{nu}}$, $M_R=10^{14}$ GeV
and the off-diagonal Dirac neutrino coupling equal to the top quark
Yukawa coupling (as expected in $SO(10)$ models), one has
$\kappa_{32}=4\times 10^{-4}$.

With this interaction, one can have a $\tau$ convert into a $\mu$ and 
a virtual $H^{0}$ or $A^{0}$, which then converts into a 
$\mu^{+}\mu^{-}$ pair.   This gives the branching ratio mentioned 
above.  However, one could equally well have an $A^{0}$ convert into a 
strange quark pair, which then becomes an $\eta$, as shown in Figure 
1, giving $\tau\ra\mu\eta$.  This will have both advantages and 
disadvantages.  The two body phase space is a major advantage, and  the
extra color factor and slightly bigger coupling (since 
$m_{s}>m_{\mu}$) are also advantages.  The disadvantage is in 
converting the strange quarks into an $\eta$.

\begin{figure}
    \centerline{ \epsfxsize 3.5in {\epsfbox{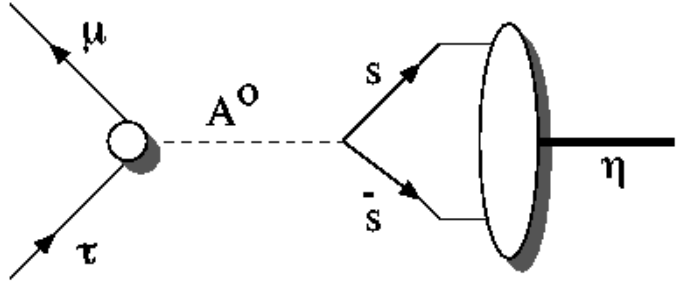}}}
    \caption{Diagram leading to $\tau\ra\mu\eta$.  The small circle is
the lepton-flavor violating vertex of Babu and Kolda.}
    \end{figure} A general discussion of $\tau\ra\mu\eta$ can be found
in Ref. 
\cite{black}.  The relevant matrix element is 
\begin{equation}
    <0|\overline{s}\gamma_{5}s|\eta> = 
    -\sqrt{6}F_{\eta}^{8}{m_{\eta}^{2}\over m_{u}+m_{d}+4m_{s}}
\end{equation}
 Using this matrix element, it is  straightforward to calculate the
branching ratio.  If one divides by  the $\tau\ra 3\mu$ branching
ratio, the unknown parameters all cancel,  and the result is
(neglecting the muon mass)
\begin{equation}
    {\Gamma(\tau\ra\mu\eta)\over \Gamma(\tau\ra 3\mu)}=
    54\pi^{2}\left({F^{8}_{\eta}\over m_{\mu}}\right)^{2}\left({
    m_{\eta}\over m_{\tau}}\right)^{4}\left( 1-{m_{\eta}^{2}\over 
    m_{\tau}^{2}}\right)^{2}
\end{equation}    With  $F^{8}_{\eta}\sim 150$ MeV\cite{f8}, this ratio
is 8.4, giving a  branching ratio of
$${\rm BR}(\tau\ra\mu\eta)= (0.84 \times
10^{-6})\times\left({\tan\beta\over  60}\right)^{6}\times \left({100\
{\rm GeV}\over m_{A}}\right)^{4}$$

One can get this result approximately without doing a calculation. 
Imagine that final state interactions are turned off and that the
strange quarks propagate as free particles.  One expects the ratio of
$\tau\ra\mu \overline{s}{s}$ to $\tau\ra 3\mu$ to have a factor of 3 for
color and a factor of $(m_s/m_\mu)^2$ for the Yukawa coupling.  The
cross diagram in the muon case turns out to lower the rate by a factor
of $3/2$, so the overall ratio is ${9m^2_s\over 2m^2_\mu}\sim 10$.  
Since the only two body decays would be $\mu\eta$ and $\mu\eta^\prime$,
and the latter is suppressed much more by phase space, the $\mu\eta$ 
rate should dominate this process.

Since the experimental bound\cite{cleo} on the branching ratio for 
$\tau\ra\mu\eta$ is $9.6\times 10^{-6}$, and that\cite{muegam} for
$\tau\ra 3\mu$  is $1.9\times 10^{-6}$, it is clear that
$\tau\ra\mu\eta$ puts  stronger constraints on the model.  In order to
reach the interesting  region of parameter space ($\tan\beta\sim 60$
and $m_{A}\sim 100$  GeV), the bound on $\tau\ra 3\mu$ would need to
be improved by a  factor of $20$, whereas the bound on $\tau\ra\mu\eta$
would need to  be improved by a factor of $10$.  

Could these improvements be made?  Both the $\tau\ra 3\mu$ and
$\tau\ra\mu\eta$ bounds are based on the CLEO-II sample of $4.7\ {\rm
fb}^{-1}$, and the CLEO experiment has now accumulated a total of 5
times that luminosity (which would give a total of 24 million tau
pairs).  So in the absence of backgrounds, the current bounds could
improve by a factor of five.   The efficiency for $\tau\ra 3\mu$ is
listed as
$15\%$; the efficiency for $\tau\ra \mu\eta$ is about $3\%$, when one
includes the fact that they only search for the $\gamma\gamma$ channel
for the $\eta$ (thus the factor of five difference in the current
bounds).  Including the three-pion decay, or increasing the fiducial
area for finding photons,  could improve that efficiency  substantially.
Over the next years, BABAR and BELLE will reach $500\ {\rm fb}^{-1}$,
which could easily reach the needed sensitivity, depending on the point
at which they become background-limited.   Even if $\tan\beta$ is
somewhat  smaller, or $m_{A}$ larger, the necessary sensitivity could
possibly  be reached at LHC, SuperKEKB or a tau-charm factory.   Note
that the
$\tau\ra 3\mu$ decay could still be dominant in the small region of
parameter space in which
$M_H << M_A$. 

Are there any other processes that can give $\tau\ra\mu\eta$?   One 
can have the box diagram of Figure 2, which will also yield $\tau$ 
decays into $\mu$ plus other mesons, including the $\pi,\rho$ and 
$\phi$.   If we take the special case in which the neutralinos are 
pure photino, the rate for $\tau\ra\mu\pi$ is given by
\begin{equation}
    {\Gamma(\tau\ra\mu\pi)\over \Gamma(\tau\ra\mu\gamma)}=
    {32\pi\alpha F_{\pi}^{2}m^{8}_{\tilde l}(I^{2}_{1}+I^{2}_{2})
    \over 81 m^{2}_{\tau} m^{8}_{\tilde\gamma} M^{2}_{3}(x)}
    \end{equation} where $x\equiv m_{\tilde\gamma}/m_{\tilde l}$,
$m_{\tilde\gamma}$ is  the photino mass, $m_{\tilde l}$ is the average
slepton mass, 
$M_{3}(x)$ is given in Ref. \cite{older}, and the integrals are 
\begin{equation}
   (I_{1},I_{2})= {1\over 16\pi^{2}}\int_{0}^{\infty} dx\ {(x,{1 \over 
    4}x^{2})\over (x-1)^{2}(x-a)(x-b)(x-c)}
    \end{equation}
    Here, $a,b,c$ are the ratios of the squark,smuon and stau masses 
    to the photino mass.  In giving this expression, we have used the 
    fact that $m_{\pi}^{2}/(m_{u}+m_{d})$ is numerically close to 
    $m_{\tau}$.    In the case of $\tau\ra\mu\eta$, there will be a
suppression of a factor of $4$ from the $s$-quark charge, an
increase of a factor of $(F_\eta^8/F_\pi)^2\sim 3.0$ from the decay
contants, and the coefficient of the $I_1^2$ term will be decreased
by a factor of $(3m^4_\eta/8m_s^2)$ relative to $m_\tau^2$, or a
factor of about $3$. This ratio has been evaluated for the entire SUSY
parameter space, assuming sparticle masses in the range of $5-1000$
GeV\cite{light}, and  is always less than
$10^{-2}$.   As a result,
$\tau\ra\mu+\ $meson  arising from the box diagram is negligible.
\begin{figure}
    \centerline{ \epsfxsize 3.5in {\epsfbox{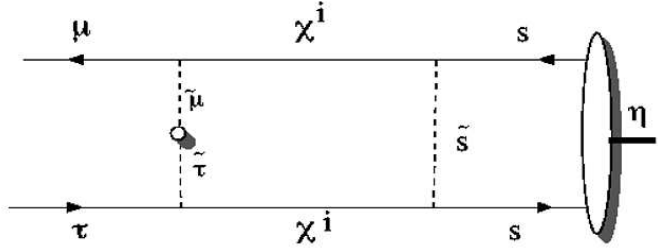}}}
    \caption{Box diagram leading to $\tau\ra\mu\eta$.  A similar 
    diagram will lead to $\tau\ra\mu\pi$.  The $\chi^i$ 
    are neutralinos.}
    \end{figure}

The existence of large $\nu_{\mu}-\nu_{\tau}$ mixing implies mixing 
between the left-handed smuon and stau.  While one could look for this 
mixing directly in neutralino/slepton interactions, one can also look 
at $\tau$ decays.  The decay $\tau\ra\mu\gamma$ is one signature, 
however Babu and Kolda have noted the mixing will also lead, 
especially in the large $\tan\beta$ region, to $\tau\ra 3\mu$.  In 
this paper, it has been pointed out that $\tau\ra\mu\eta$ will also 
occur in this Higgs-mediated model, with a branching ratio $8.4$ times 
bigger, and is thus more sensitive.  In other models, where 
$\tau\ra\mu\gamma$ is the main signature, the $\tau\ra\mu\eta$ rate  is
substantially smaller.

I thank Carl Carlson, Chris Carone and Jon Urheim for useful
discussions.

\end{document}